\documentclass[12pt, preprint]{aastex}
\usepackage{amsmath}
\usepackage{graphicx}
\usepackage{natbib}
\usepackage{ulem}

\usepackage[colorlinks=true,citecolor=blue,breaklinks=true,linktocpage=true]{hyperref}
\bibpunct{(}{)}{;}{a}{}{,} %% natbib format like A&A and ApJ
\usepackage{xspace}
\usepackage{hyperref}

\usepackage{multirow}

\def\vec#1{\ensuremath{\mathbf{#1}}}

\usepackage{color}

\newcommand{\omits}[1]{}

\shorttitle{Fast modes in nonuniform coronal slabs with finite plasma beta}
\shortauthors{Chen et al.}

\begin{document}


\title{FAST STANDING MODES IN TRANSVERSELY NONUNIFORM SOLAR CORONAL SLABS: EFFECTS OF A FINITE PLASMA BETA}

\author{Shao-Xia Chen\altaffilmark{1}}
\author{Bo Li\altaffilmark{1}}
    \email{bbl@sdu.edu.cn}
\author{Sanjay Kumar\altaffilmark{1}}
\author{Hui Yu\altaffilmark{1}}
\and
\author{Mijie Shi\altaffilmark{1}}

\altaffiltext{1}{Shandong Provincial Key Laboratory of Optical Astronomy and Solar-Terrestrial Environment,
   Institute of Space Sciences, Shandong University, Weihai 264209, China}

\begin{abstract}
We examine the dispersive properties of linear fast standing modes in transversely nonuniform solar coronal slabs with finite gas pressure, or, equivalently, finite plasma beta. We derive a generic dispersion relation governing fast waves in coronal slabs for which the continuous transverse distributions of the physical parameters comprise a uniform core, a uniform external medium, and a transition layer (TL) in between. The profiles in the TL are allowed to be essentially arbitrary. Restricting ourselves to the first several branches of fast modes, which are of most interest from the observational standpoint, we find that a finite plasma beta plays an at most marginal role in influencing the periods ($P$), damping times ($\tau$), and critical longitudinal wavenumbers ($k_{\rm c}$), when both $P$ and $\tau$ are measured in units of the transverse fast time. However, these parameters are in general significantly affected by how the TL profiles are described. We conclude that, for typical coronal structures, the dispersive properties of the first several branches of fast standing modes can be evaluated with the much simpler theory for cold slabs provided that the transverse profiles are properly addressed and the transverse Alfv\'en time in cold MHD is replaced with the transverse fast time.
\end{abstract}

\section{INTRODUCTION}
\label{sec_intro}

The past two decades have seen rapid development of solar magneto-seismology (SMS) in general,
    and coronal seismology in particular.
On the one hand, this was made possible by the abundant measurements of low-frequency waves and oscillations
    in the Sun's atmosphere
    \citep[see, e.g.,][for recent reviews]{2007SoPh..246....3B, 2012RSPTA.370.3193D,2016GMS...216..395W}.
On the other hand, an increasingly refined theoretical understanding
    has become available
    regarding the collective wave modes in magnetized plasma inhomogeneities
    \citep[see, e.g.,][]{2000SoPh..193..139R, 2005LRSP....2....3N, 2008IAUS..247....3R, 2016SSRv..200...75N}.
For this latter purpose, plasma inhomogeneities are traditionally seen as field-aligned magnetic cylinders,
    following the seminal studies by
    e.g.,
    \citeauthor{1970A&A.....9..159R}~\citeyear{1970A&A.....9..159R},
    \citeauthor{1975IGAFS..37....3Z}~\citeyear{1975IGAFS..37....3Z},
    \citeauthor{1982SoPh...75....3S}~\citeyear{1982SoPh...75....3S},
    \citeauthor{1983SoPh...88..179E}~\citeyear{1983SoPh...88..179E},
    and
    \citeauthor{1986SoPh..103..277C}~\citeyear{1986SoPh..103..277C}.
In fact, the modern terminology in SMS has largely come from this series of studies, allowing one
    to distinguish between body and surface modes
    or leaky and trapped modes~\citep[see, e.g.,][]{2005LRSP....2....3N}.
Among the infinitely many modes that a plasma cylinder can host,
    kink (with azimuthal wavenumber $m=1$) and sausage ($m=0$) modes have received most attention.
This is understandable given that kink modes are the only ones that displace the cylinder axis
    and therefore may naturally account for the extensively measured transverse displacements of
    magnetic flux tubes in both the corona~\citep[e.g.,][]{1999ApJ...520..880A,1999Sci...285..862N,2008A&A...482L...9O,2008A&A...489L..49E,2009ApJ...698..397V,
    2011ApJ...736..102A} and the chromosphere~\citep[e.g.,][]{2012NatCo...3E1315M,2013ApJ...779...82K,2017ApJS..229....9J,2017arXiv170803500M}.
On the other hand, sausage modes in flare loops have been invoked to account for a significant fraction
    of quasi-periodic pulsations in solar flare light curves
    \citep[see, e.g.,][for recent reviews]{2009SSRv..149..119N,2016SoPh..291.3143V}.
There have also been a substantial number of pieces of observational evidence showing the
    existence of sausage modes in magnetic pores~\citep[e.g.,][]{2008IAUS..247..351D,2011ApJ...729L..18M,2014A&A...563A..12D,2015ApJ...806..132G,2016ApJ...817...44F}
    and coronal loops~\citep[e.g.,][]{2012ApJ...755..113S,2016ApJ...823L..16T,2017ApJ...836...84D}.

Wave-guiding plasma inhomogeneities in the solar atmosphere
    have also been extensively modeled as magnetic slabs
    \citep[e.g.,][]{1982SoPh...76..239E,1993SoPh..143...89M,1994ApJ...421..360O,1994ApJ...435..502P,1999A&A...342..300H,2005A&A...441..371T,2013ApJ...767..169L,
    2014SoPh..289.1663C, 2014A&A...567A..24H, 2015ApJ...814...60Y,2016ApJ...826...78Y,2017SoPh..292...35A}.
This practice can be justified given that a slab rather than a cylindrical geometry
    is more suitable for interpreting a considerable number of observations.
For instance, Sunward moving tadpole-like structures in post-flare supra-arcades as observed by
    the Transition Region And Coronal Explorer (TRACE) were attributed to fast kink waves in vertical slab-like structures
    \citep{2005A&A...430L..65V}.
The extensively measured prominence oscillations were also
    interpreted in terms of collective modes in dense slabs embedded in the corona
    \citep[see][and references therein]{2012LRSP....9....2A}.
Interestingly, the theoretical results on collective modes in wave-guiding slabs also apply
    even when current sheets are embedded, albeit in the limit
    of vanishing resistivity~\citep[e.g.,][]{1986GeoRL..13..373E, 1997A&A...327..377S, 2011SoPh..272..119F}.
It is then no surprise to see that
    fast kink waves in magnetic slabs were invoked to account for
    the large-scale propagating disturbances
    in current sheets above streamer helmets, as imaged
    with the Large Angle and Spectrometric Coronagraph (LASCO) on board the Solar and Heliospheric
    Observatory satellite (SOHO)~\citep{2010ApJ...714..644C,2011ApJ...728..147C}.
Furthermore, fast sausage modes guided by flare current sheets were employed to interpret
    some fine structures in broadband type IV radio bursts~\citep[e.g.,][]{2012A&A...537A..46J,
	2013A&A...550A...1K,2016ApJ...826...78Y}.

For mathematical simplicity, analytical studies of fast
    modes in magnetic slabs usually adopt either one or both of the following two assumptions.
One is the cold (zero-$\beta$) MHD limit, where thermal pressure is neglected.
The other is that the plasma and magnetic parameters are transversely structured in a top-hat fashion.
Relaxing the second assumption,
    some studies are available where the continuous transverse profiles are
    either some specific ones~\citep[e.g.,][]{1988A&A...192..343E,1995SoPh..159..399N,2011A&A...526A..75M, 2015ApJ...801...23L}
    or essentially arbitrary~\citep[][hereafter paper I]{2015ApJ...814...60Y}.
Relaxing the first assumption,
    available studies nonetheless tend to adopt a top-hat profile
    for the transverse distributions of physical parameters
    \citep[e.g.,][]{1982SoPh...76..239E,2009A&A...503..569I}.
In cylindrical geometry, we have offered an analytical study on
    fast sausage modes in coronal tubes for which the gas pressure is finite
    and the transverse profiles are essentially arbitrary
    \citep[][hereafter paper II]{2016ApJ...833..114C}.
The aim of the present manuscript is to offer a slab counterpart of our paper II.

Before proceeding, a few words seem necessary to justify the motivation of this study.
First, such parameters as temperature, density, and the magnetic field strength
    are evidently more likely to be continuously distributed across magnetic structures.
Second, the plasma $\beta$, which measures the importance of the gas pressure relative
    to the magnetic one, is not necessarily small
    but may reach a value of order unity in hot and dense solar structures \citep[e.g.,][]{2005A&A...439..727M,2007ApJ...656..598W}.
Third, while the combined effects of continuous transverse profiles
    and finite gas pressure can be readily addressed by numerical simulations
    \citep[e.g.,][]{2009A&A...503..569I,2012A&A...537A..46J,2016ApJ...826...78Y},
    performing a largely analytical eigen-mode analysis for fast collective modes still proves useful.
On the one hand, examining the dispersive properties is much less numerically expensive.
On the other hand, the frequency-dependence of the axial group speed can be readily computed,
    which will facilitate the interpretation of the numerical results on
    impulsively generated wave trains
    in, say, flare current sheets~\citep{2012A&A...546A..49J,2013A&A...550A...1K,2014ApJ...788...44M}.
The applications in this regard have been offered in
    \citet{2016ApJ...833...51Y}
    and \citet{2017ApJ...836....1Y}.

This manuscript is structured as follows.
Section \ref{sec_equilibrium} presents the necessary description
    for an equilibrium straight magnetic slab.
In Section \ref{sec_DR}, we derive a generic dispersion relation (DR)
    for linear magneto-acoustic waves in magnetic slabs
    with nonneglible gas pressure and rather arbitrary transverse distributions.
We examine, in Section \ref{sec_para_study},
     the effects of a finite $\beta$ on the dispersive properties of standing fast modes in substantial detail.
Section \ref{sec_conclusion} closes this manuscript with our summary
    and some concluding remarks.

\section{DESCRIPTION FOR THE EQUILIBRIUM SLAB}
\label{sec_equilibrium}

\subsection{Overall Description}
\label{sec_sub_equil_overall}

We model coronal structures as
    straight, gravity-free, density-enhanced slabs with mean half-width $R$.
We adopt a Cartesian coordinate system $(x,~y,~z)$,
    where the $z$-axis coincides with the axis of the slab.
The equilibrium magnetic field $\vec{B}$ is
    in the $z$-direction and is a function of $x$ only.
We further assume that both plasma density ${\rho}$ and
    temperature $T$ depend only on $x$.
Note that we are allowed to prescribe the $x$-dependence of
    only two parameters in $[B, \rho, T]$, since they
    are related by the transverse force balance condition
\begin{equation}
      {p}(x)+\displaystyle\frac{{B}^2(x)}{8\pi}= \mbox{const} \equiv \alpha~,
\label{eq_force_balance}
\end{equation}
    in which the gas pressure $p$ is given by
\begin{equation}
  {p} = \displaystyle\frac{2 k_{\rm B}}{m_p}  {\rho} {T}.
\label{eq_state}
\end{equation}
Here $k_{\rm B}$ is the Boltzmann constant and $m_p$ the proton mass.

The following characteristic speeds are necessary for
    us to proceed.
The adiabatic sound and Alfv\'en
    speeds are defined as
 \begin{equation}
   c_{\rm s}^2=\displaystyle\frac{\gamma  {p}}{ {\rho}},
   \hspace{0.5cm} \mbox{and} \hspace{0.5cm}
   v^2_{\rm A}=\displaystyle\frac{ {B}^2}{4\pi {\rho}},
   \label{eq_def_cs_va}
 \end{equation}
     where $\gamma=5/3$ is the adiabatic index.
It then follows that the plasma $\beta \equiv 8\pi  {p}/ {B}^2$ reads
\begin{equation}
   \beta = \frac{2}{\gamma}\frac{c^2_{\rm s}}{v^2_{\rm A}} .
   \label{eq_def_beta}
\end{equation}
We then define the fast and tube speeds as
 \begin{equation}
   v_{\rm f}^2= c_{\rm s}^2 +v^2_{\rm A},
   \label{eq_def_vf}
 \end{equation}
    and
\begin{equation}
    c_{\rm T}^2=\displaystyle\frac{c_{\rm s}^2v^2_{\rm A}}{c_{\rm s}^2+v^2_{\rm A}},
    \label{eq_def_ct}
\end{equation}
    respectively.

\subsection{Description for Transverse Profiles}
We assume that the equilibrium parameters are symmetric about $x=0$.
In the half-plane $x>0$, we choose to
    prescribe the equilibrium density $ {\rho}(x)$
    and temperature $ {T}(x)$ as
\begin{eqnarray}
  {\rho}(x)=\left\{
   \begin{array}{ll}
   \rho_{\rm i},    							& 0\le x \leq x_{\rm i} = R-l/2 ,\\
   \rho_{\rm tr}(x) = {\cal F}(\rho_{\rm i}, \rho_{\rm e}; x), 		& x_{\rm i} \le x \le x_{\rm e} = R+l/2, \\ \label{eq_profile_rho_overall}
   \rho_{\rm e},    							& x \ge x_{\rm e} ,
   \end{array}
   \right.
\end{eqnarray}
    and
\begin{eqnarray}
  {T}(x)=\left\{
   \begin{array}{ll}
   T_{\rm i},    				& 0\le x \leq x_{\rm i} , \\
   T_{\rm tr}(x) = {\cal G}(T_{\rm i}, T_{\rm e}; x),	& x_{\rm i} \le x \le x_{\rm e} ,\\ \label{eq_profile_T_overall}
   T_{\rm e},    				& x \ge x_{\rm e}.
   \end{array}
   \right.
\end{eqnarray}
This sandwich-like profile means that the equilibrium configuration comprises a uniform core
    (denoted
    by subscript ${\rm i}$), a uniform external medium (subscript ${\rm e}$),
    and a transition layer (TL) connecting the two.
This TL is of width $l$ and centered around the mean
    half-width $R$.
Furthermore, ${\cal F}(\rho_{\rm i}, \rho_{\rm e}; x)$
    and ${\cal G}(T_{\rm i}, T_{\rm e}; x)$ are functions that continuously vary from
    the core-TL interface ($x=x_{\rm i}$)
    to the TL-external-medium interface ($x=x_{\rm e}$).
In addition, 
    we require that the functions ${\cal F}$ and ${\cal G}$ are smooth at $x=R$,
    making it possible to Taylor expand $\rho_{\rm tr}(x)$
    and $T_{\rm tr}(x)$ around $x = R$.
The result is
\begin{eqnarray}
  \rho_{\rm tr}(\epsilon) = \sum^\infty_{n=0}\rho_n \epsilon^n,~~~~~~~~T_{\rm tr}(\epsilon) = \sum^\infty_{n=0}T_n \epsilon^n ,
\label{eq_rhoT_expansion}
\end{eqnarray}
    where $\epsilon\equiv x-R$, $\rho_0 =  {\rho}|_{\epsilon=0}$, $T_0 =  {T}|_{\epsilon=0}$ and
\begin{eqnarray}
\label{eq_rho_coef}
  \rho_n = \frac{1}{n!} \left.\frac{{\rm d}^n {\rho}(\epsilon)}{{\rm d}\epsilon^n}\right|_{\epsilon=0},~~~~~~~~~~~T_n = \frac{1}{n!} \left.\frac{{\rm d}^n {T}(\epsilon)}{{\rm d}\epsilon^n}\right|_{\epsilon=0},~~~~~~~~~~~
     \hspace{0.2cm} n\ge 1 .
\end{eqnarray}
In the TL, we Taylor-expand the characteristic speeds
    $c_{\rm s}^2$ and $v_{\rm A}^2$ as
\begin{eqnarray*}
    c_{\rm s}^2 = \sum\limits^\infty_{n=0}C_n \epsilon^n, \hspace{0.5cm}
    v_{\rm A}^2 = \sum\limits^{\infty}_{n=0}V_n \epsilon^n~,
\end{eqnarray*}
    where
\begin{equation}
  C_n=\displaystyle\frac{2\gamma k_{\rm B}}{m_p}T_n ,
  \label{eq_coef_Cn}
\end{equation}
    and
\begin{equation}
  \left\{
  \begin{array}{rcl}
  V_0 &=& \displaystyle\frac{2\alpha}{\rho_0}-
          \displaystyle\frac{2}{\gamma}C_0 ,\\ [0.3cm]
  V_n &=& -\displaystyle\frac{1}{\rho_0}\left(
           \displaystyle\frac{2}{\gamma}\sum\limits^{n}_{l=0}C_l \rho_{n-l}+\sum\limits^{n-1}_{l=0}V_l \rho_{n-l}
  \right),~~~~~~n\ge1 .
  \end{array}
  \right.
  \label{eq_coef_Vn}
 \end{equation}
Note that the force balance condition (Eq.~\ref{eq_force_balance}) has been used
    to derive the coefficients $V_n$.

\section{DISPERSION RELATIONS OF FAST WAVES}
\label{sec_DR}
\subsection{Dispersion Relations for Arbitrary Transverse Profiles in the TL}
Now we examine linear fast waves,
    by which we mean waves with phase speeds always exceeding $c_s$, 
    in magnetic slabs pertaining to the configuration
    that was just described.
For this purpose we adopt the framework of ideal MHD.
Let $\delta \rho$, $\delta \vec{v}$, $\delta \vec{B}$,
    and $\delta p$ denote
    the perturbations to the density, velocity, magnetic field and pressure, respectively.
We consider only fast waves in the $x-z$ plane
    by letting $\partial/\partial y\equiv0$, and $\delta B_y = \delta v_y =0$.
Any perturbation $\delta f(x, z;t)$  can be Fourier-decomposed as
\begin{eqnarray}
\label{eq_Fourier_ansatz}
  \delta f(x,z;t)={\rm Re}\left\{\tilde{f}(x)\exp\left[-i\left(\omega t-kz\right)\right]\right\}~.
\end{eqnarray}
Now with the definition of the Fourier amplitude
    for the Lagrangian displacement $\tilde{\xi}_x = i\tilde{v}_x/\omega$,
    it is straightforward to show that $\tilde{\xi}_x$ is governed by
\begin{equation}
  \displaystyle\frac{{\rm d}}{{\rm d}x}\left[\displaystyle\frac{ {\rho}(c_{\rm s}^2+v^2_{\rm A})(\omega^2-k^2c_{\rm T}^2)}{\omega^2-k^2c_{\rm s}^2}
  \displaystyle\frac{{\rm d}\tilde{\xi}_x}{{\rm d}x}\right]+ {\rho}(\omega^2-k^2v_{\rm A}^2)\tilde{\xi}_x=0.
\label{eq_govern_y}
\end{equation}
The solution to Equation (\ref{eq_govern_y}) in a
    uniform medium is well-known \citep[e.g.,][]{1982SoPh...76..239E}.
With coronal applications in mind,
    we assume that the ordering $v_{\rm Ae} > v_{\rm Ai} > c_{\rm si} > c_{\rm se}$ holds and
    $c_{\rm s}<v_{\rm A}$ in the TL.
In this case, Equation (\ref{eq_govern_y})
    is singularity-free for fast waves.
This is because fast modes always propagate faster than the tube speed, and hence cannot couple
    to the slow continuum.
Furthermore, when propagating strictly in the $x-z$ plane,
    they are not resonantly coupled to the shear Alfv\'en continuum either~\citep[e.g.,][]{2007SoPh..246..213A}.
The solution to Equation (\ref{eq_govern_y}) in the TL
    can then be expressed as linear combinations of two linearly independent solutions,
    $\tilde\xi_{\rm tr,1}$ and $\tilde\xi_{\rm tr,2}$,
\begin{eqnarray}
    \tilde\xi_{\rm tr,1}(\epsilon) = \sum_{n=0}^\infty a_n \epsilon^n~, \hspace{0.2cm}
    \tilde\xi_{\rm tr,2}(\epsilon) = \sum_{n=0}^\infty b_n \epsilon^n~.
\label{eq_def_y1y2_expansion}
\end{eqnarray}

Now the standard practice will be to insert the regular series expansion (\ref{eq_def_y1y2_expansion})
    into Equation (\ref{eq_govern_y}) such that the recurrence relations for the coefficients $a_n$ and $b_n$ can be found.
For this purpose, however, it turns out to be more convenient to work with an equation where $\rho$ is not directly present.
Given the force balance condition (\ref{eq_force_balance}), it is straightforward to show that $\rho$ can be expressed as
\begin{equation}
  \rho = \frac{2\alpha}{v_{\rm A}^2+2c_{\rm s}^2/\gamma}~. \nonumber
\end{equation}
As a result, Equation (\ref{eq_govern_y}) can be reformulated into
\begin{equation}\label{eq_govern_new}
 \displaystyle\frac{{\rm d}}{{\rm d}x}\left[\displaystyle\frac{(c_{\rm s}^2+v^2_{\rm A})(\omega^2-k^2c_{\rm T}^2)}{(2c_{\rm s}^2+\gamma v^2_{\rm A})\left(\omega^2-k^2c_{\rm s}^2\right)}
  \displaystyle\frac{{\rm d}\tilde{\xi}_x}{{\rm d}x}\right]+\displaystyle\frac{\omega^2-k^2v_{\rm A}^2}
  {2c_{\rm s}^2+\gamma v^2_{\rm A}}\tilde{\xi}_x=0.
\end{equation}
Then the recurrence relations can be found by inserting the expansion (\ref{eq_def_y1y2_expansion}) into
    Equation (\ref{eq_govern_new}) and demanding the coefficient of $\epsilon^n~(n=0,~1,~2,~\cdots)$ to be zero.
Without loss of generality, we choose
\begin{equation}
  a_0 = R,~a_1 = 0,~\mbox{and}~b_0 = 0,~b_1 =1 .
\label{eq_def_a01_b01}
\end{equation}
The expressions for the rest of the coefficients are too lengthy and therefore
    are given in Appendix \ref{sec_app_coef_general}.
It suffices to note that they involve only the coefficients $C_n$ and $V_n$.
All in all, $\tilde{\xi}_x(x)$ can be expressed as
\begin{eqnarray}
   \tilde{\xi}_x(x)=\left\{
   \begin{array}{ll}
      \left\{
      \begin{array}{cc}
      A_{\rm i}\sin(\mu_{\rm i}x) & {\rm sausage} \\ [0.2cm]
      A_{\rm i}\cos(\mu_{\rm i}x) & {\rm kink}
      \end{array}
      \right.,          & 0         \le x \le x_{\rm i}, \\ [0.5cm]
      A_1\tilde\xi_{\rm tr,1}(\epsilon)+A_2\tilde\xi_{\rm tr,2}(\epsilon),      & x_{\rm i} \le x \le x_{\rm e}, \\ [0.2cm]
      A_{\rm e}\exp\left(i\mu_{\rm e}x\right),        & x \ge x_{\rm e},
   \end{array} \right.
\label{eq_y_solution_entire}
\end{eqnarray}
    where $A_{\rm i},~A_{\rm e},~A_1$ and $A_2$ are arbitrary constants.
In addition,
\begin{equation}
 \mu_{\rm i, e}^2 =\displaystyle\frac{(\omega^2 - k^2v_{\rm Ai, e}^2)(\omega^2 - k^2c_{\rm si, e}^2)}{(c_{\rm si, e}^2+v_{\rm Ai, e}^2)(\omega^2 - k^2c_{\rm Ti, e}^2)}~.
\label{eq_def_mu}
\end{equation}

To derive the DR, one also requires the explicit expressions
    for the Fourier amplitude of the Eulerian perturbation
    of total pressure $\tilde{p}_{\rm T}$.
It is related to the Lagrangian displacement via \citep[e.g.,][]{1982SoPh...76..239E}
\begin{eqnarray}
  \tilde{p}_{\rm T} = -
     \displaystyle\frac{ {\rho}(c_{\rm s}^2+v_{\rm A}^2)(\omega^2 - k^2c_{\rm T}^2)}{\omega^2 - k^2c_{\rm s}^2}\tilde{\xi}_x' ,
\label{eq_Fourie_ptot_xi}
\end{eqnarray}
   where the prime $' = {\rm d}/{\rm d} x$.
With the aid of Equation (\ref{eq_y_solution_entire}), one finds that
\begin{equation*}
\tilde{p}_{\rm T}(x)=
  \left\{ \begin{array}{ll}
    -A_{\rm i}\mu_{\rm i}
     \displaystyle\frac{\rho_{\rm i}(c_{\rm si}^2+v_{\rm Ai}^2)(\omega^2 - k^2c_{\rm Ti}^2)}{\omega^2 - k^2c_{\rm si}^2}\left\{
      \begin{array}{cc}
      \cos(\mu_{\rm i}x) & {\rm sausage} \\ [0.2cm]
      -\sin(\mu_{\rm i}x) & {\rm kink}
      \end{array}
      \right.,
        & 0 \le x \leq x_{\rm i}, \\[0.7cm]
    -iA_{\rm e}\mu_{\rm e}\exp\left(i\mu_{\rm e}x\right)\displaystyle\frac{\rho_{\rm e}(c_{\rm se}^2+v_{\rm Ae}^2)(\omega^2 - k^2c_{\rm Te}^2)}{\omega^2 - k^2c_{\rm se}^2}~,    & x \ge x_{\rm e},
   \end{array}
   \right.
\end{equation*}
    in the uniform core and external medium.
In the TL it is given by
\begin{eqnarray}
   \tilde{p}_{\rm T}(\epsilon)= -
     \displaystyle\frac{\sum\limits^\infty_{l=0}\rho_l \epsilon^l\left(\omega^2\sum\limits^\infty_{n=0}C_n \epsilon^n+\omega^2\sum\limits^\infty_{n=0}V_n \epsilon^n - k^2\sum\limits^\infty_{n=0}C_n \epsilon^n\sum\limits^\infty_{j=0}V_j \epsilon^j\right)}
     {\omega^2 - k^2\sum\limits^\infty_{n=0}C_n\epsilon^n}
        \left[ A_1\tilde\xi_{\rm tr,1}'(\epsilon)
               +A_2\tilde\xi_{\rm tr,2}'(\epsilon)
        \right].
\end{eqnarray}
Requiring that $\tilde{\xi}_x$ and $\tilde{p}_{\rm T}$ be continuous
     at $x=x_{\rm i}$ and $x=x_{\rm e}$ yields four algebraic equations
     governing $[A_1,~A_2,~A_{\rm i},~A_{\rm e}]$.
For these solutions to be non-trivial, one finds that
\begin{equation}
 \displaystyle\frac{
  X_{\rm i}\tilde\xi_{\rm tr,1}(\epsilon_{\rm i})+\Lambda_{\rm i}\tilde\xi_{\rm tr,1}'(\epsilon_{\rm i})
  }
  {
  X_{\rm i}\tilde\xi_{\rm tr,2}(\epsilon_{\rm i})+\Lambda_{\rm i}\tilde\xi_{\rm tr,2}'(\epsilon_{\rm i})
  }
  -\displaystyle\frac{
  X_{\rm e}\tilde\xi_{\rm tr,1}(\epsilon_{\rm e})+\Lambda_{\rm e}\tilde\xi_{\rm tr,1}'(\epsilon_{\rm e})
  }
  {
  X_{\rm e}\tilde\xi_{\rm tr,2}(\epsilon_{\rm e})+\Lambda_{\rm e}\tilde\xi_{\rm tr,2}'(\epsilon_{\rm e})
  }
  =0~,
\label{eq_DR}
\end{equation}
    in which $\epsilon_{\rm i,e} = \mp l/2$ and
\begin{equation}
\begin{array}{ll}
 &\Lambda_{\rm i,e}=-
\displaystyle\frac{\sum\limits^\infty_{l=0}\rho_l \epsilon^l_{\rm i,e}\left(\omega^2\sum\limits^\infty_{n=0}C_n \epsilon^n_{\rm i,e}+\omega^2\sum\limits^\infty_{n=0}V_n \epsilon^n_{\rm i,e} - k^2\sum\limits^\infty_{n=0}C_n\epsilon^n_{\rm i,e}\sum\limits^\infty_{j=0}V_j\epsilon^j_{\rm i,e}\right)}
     {\omega^2 - k^2\sum\limits^\infty_{n=0}C_n\epsilon_{\rm i,e}^n}~,\\ [0.8cm]
     & X_{\rm i}=\displaystyle\frac{\rho_{\rm i}\left(\omega^2-k^2v_{\rm Ai}^2\right)}{\mu_{\rm i}}\times
     \left\{
      \begin{array}{cc}
      \cot(\mu_{\rm i}x_{\rm i}) & {\rm sausage} \\ [0.2cm]
      -\tan(\mu_{\rm i}x_{\rm i}) & {\rm kink}
      \end{array}
      \right.
     ~,~~~~~~~~~~X_{\rm e}=\displaystyle\frac{i\rho_{\rm e}\left(\omega^2-k^2v_{\rm Ae}^2\right)}{\mu_{\rm e}}~.
 \label{eq_Lambda}
 \end{array}
 \end{equation}
Equation~(\ref{eq_DR}) is the DR we are looking for, and is valid for rather arbitrary choices
    of the transverse profiles in the TL.

\subsection{Dispersion Relation for Top-hat Transverse Profiles}
\label{sec_sub_tophat}
For future reference, this section examines what happens for top-hat profiles by letting $l/R \rightarrow 0$
   in the DR~(\ref{eq_DR}).
To do this we retain only terms to the zeroth order in $l/R$.
Hence $x_{\rm i}\approx x_{\rm e} \approx R$ and $\epsilon_{\rm i}\approx \epsilon_{\rm e}\approx 0$.
Now that $\Lambda_{\rm i}\approx\Lambda_{\rm e}$,
     it follows from Equation (\ref{eq_DR}) that
\begin{eqnarray}
\label{top_X}
\left(X_{\rm i}-X_{\rm e}\right)(a_1 b_0-a_0 b_1)=0 .
\end{eqnarray}
The term $a_1 b_0 - a_0 b_1$ is not allowed to be zero
    because $\tilde\xi_{\rm tr,1}(\epsilon)$ and $\tilde\xi_{\rm tr,2}(\epsilon)$ are linearly independent.
As a result, $X_{\rm i}=X_{\rm e}$.
Recalling the definitions for $X_{\rm i}$ and $X_{\rm e}$ as given in Equation (\ref{eq_Lambda}),
    the DR in this limit simplifies to the well-known form~\citep[e.g.,][]{1982SoPh...76..239E}
\begin{eqnarray}
\displaystyle\frac{i\rho_{\rm e}\left(\omega^2-k^2v_{\rm Ae}^2\right)}{\mu_{\rm e}}=\displaystyle\frac{\rho_{\rm i}\left(\omega^2-k^2v_{\rm Ai}^2\right)}{\mu_{\rm i}}
     \times
     \left\{
      \begin{array}{cc}
      \cot(\mu_{\rm i}R) & {\rm sausage} \\ [0.2cm]
      -\tan(\mu_{\rm i}R) & {\rm kink}
      \end{array}
      \right.
     ~.
\label{eq_DR_tophat}
\end{eqnarray}

For kink and sausage modes alike,
    there are infinitely many branches of solutions to Equation~(\ref{eq_DR_tophat}),
    with the $n$-th ($n=1, 2, 3, \cdots$)
    branch characterized by the appearance of $n-1$ permanent transverse nodes for $\tilde{\xi}_x$ in the interval $0 < x <\infty$.
Except for the first kink branch,
    there always exists a critical wavenumber $k_{\rm c}$ below which fast waves are no longer trapped.
This $k_{\rm c}$ can be expressed as~\citep[e.g.,][]{1995SoPh..159..213N,2013ApJ...767..169L}
\begin{equation}\label{eq_kc_tophat}
  k_{\rm c}R=g_n\displaystyle\sqrt{\frac{\left(c_{\rm si}^2+v_{\rm Ai}^2\right)\left(v_{\rm Ae}^2-c_{\rm Ti}^2\right)}{\left(v_{\rm Ae}^2-c_{\rm si}^2\right)\left(v_{\rm Ae}^2-v_{\rm Ai}^2\right)}}~
\end{equation}
    with $g_n$ given by
    \begin{equation}\label{eq_gl}
    g_n=\left\{
      \begin{array}{cc}
      \left(n-\displaystyle\frac{1}{2}\right)\pi & {\rm sausage}~, \\ [0.4cm]
      (n-1)\pi & {\rm kink}~,
      \end{array}
      \right.
    \end{equation}
    where $n=1,~2,~3,\cdots$.
In addition, one can readily derive the expressions for the angular frequencies $\omega$
    for fast waves in the long-wavelength (or thin-slab) limit by letting
    $k R \rightarrow 0$.
In this case,
    the angular frequency for the first kink branch
    approaches zero.
For the rest of the fast modes, $\omega$ can be shown to have the following form
\begin{equation}\label{eq_omega0}
\begin{array}{rcl}
  \left. \omega_{\rm R} \right|_{k\rightarrow 0} &=&g_n\displaystyle\frac{v_{\rm fi}}{R}~,\\[0.4cm]
  \left. \omega_{\rm I} \right|_{k\rightarrow 0} &=&
     -\displaystyle\frac{v_{\rm fi}}{2R}\ln\displaystyle\frac{1+\rho_{\rm e}v_{\rm fe}/(\rho_{\rm i}v_{\rm fi})}{1-\rho_{\rm e}v_{\rm fe}/(\rho_{\rm i}v_{\rm fi})}~,
\end{array}
\end{equation}
    where $v_{\rm fi,e}$ is the internal (external) fast speed.
In addition, $\omega_{\rm R}$ ($\omega_{\rm I}$) represents the real (imaginary) part, with the non-zero $\omega_{\rm I}$
    arising from lateral leakage.
(Note that $g_n$ is different for kink and sausage modes.)
Equation~(\ref{eq_omega0}) extends the results
    in \cite{2005A&A...441..371T} by allowing for a finite plasma beta.
Two things are immediately clear.
One is that the damping time associated with lateral leakage is
    independent of the branch label $n$.
The other is that the angular frequency is associated with the transverse fast
    time $R/v_{\rm fi}$.
This is understandable given that in the long-wavelength limit, the transverse spatial scale
    of the eigen-functions is much shorter than the longitudinal wavelength, thereby
    making the effective wavevector largely perpendicular to the equilibrium magnetic field.
Hence the transverse fast time becomes relevant, even though, strictly speaking,
    the fast speed $v_{\rm f}$ as defined by Eq.~(\ref{eq_def_vf}) pertains to
    perpendicularly propagating fast waves in a uniform MHD medium.

\section{NUMERICAL RESULTS}
\label{sec_para_study}

\subsection{Prescriptions for Transition Layer Profiles and Method of Solution}
\label{sec_sub_solmethod}

\begin{figure}
\centering
 \includegraphics[width=1.\columnwidth]{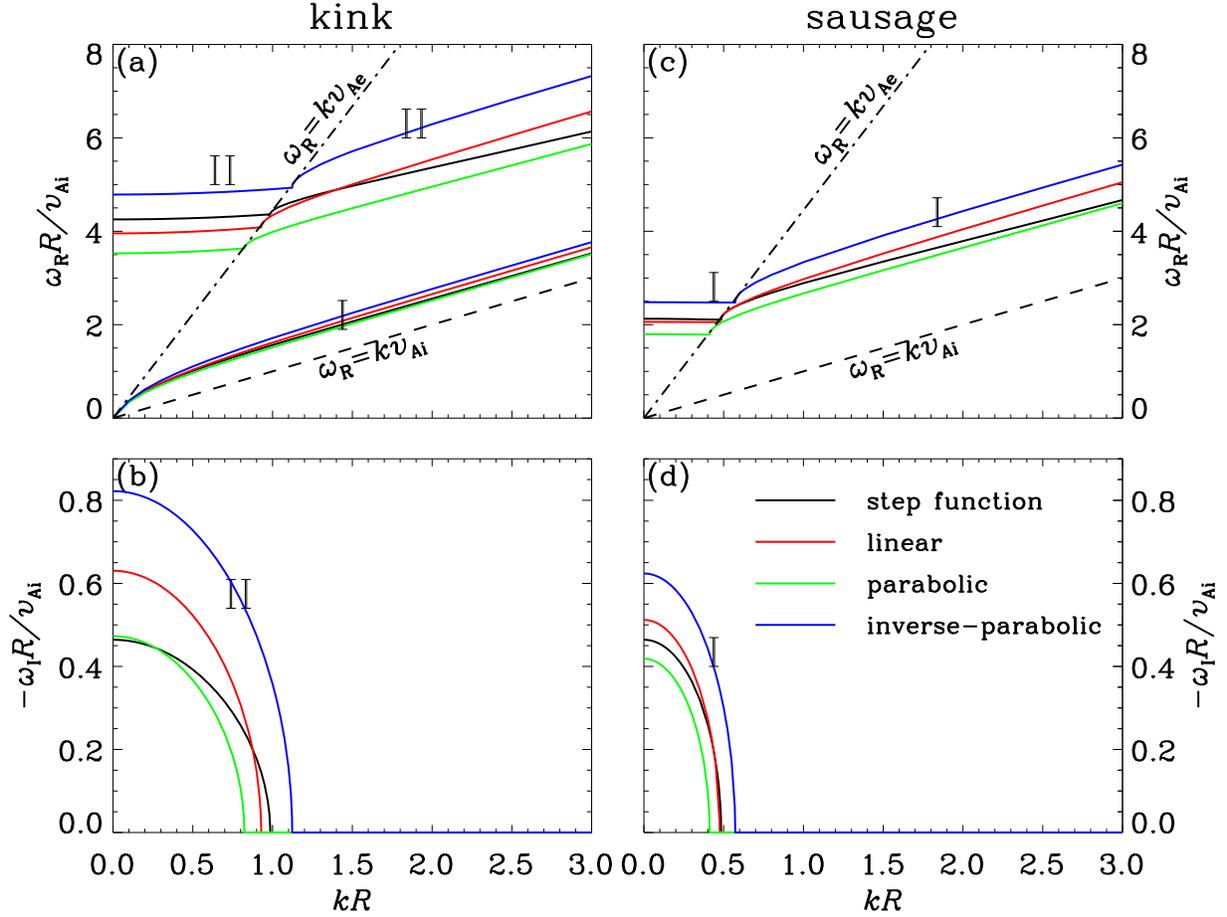}
 \caption{Illustration of profile prescriptions using
    transverse equilibrium density profiles as an example.
 These profiles differ only in a transition layer
    sandwiched between the internal (with a uniform density $\rho_{\rm i}$) and external (with a uniform density $\rho_{\rm e}$) portions.
 This transition layer is of width $l$ and is located
    between $x_{\rm  i} = R-l/2$ and $x_{\rm  e} = R+l/2$, with $R$ being the mean half-width.
 Three different profile prescriptions are
    adopted as labeled and are given by Equation (\ref{eq_TL_profile}).
 For illustration purposes, 
    {$l=R$ and $\rho_{\rm i}/\rho_{\rm e}=50$}.
}
 \label{fig_profile}
\end{figure}

Except for the assumption that $c_{\rm s}(x) < v_{\rm A}(x)$,
    no restrictions were imposed on
    the profiles $ {T}(x)$ and $ {\rho}(x)$ in the TL
    when the DR (\ref{eq_DR}) was derived.
However, the transcendental nature of the DR means that
    in general it needs to be solved numerically.
To this end, the profiles
    for $ {T}(x)$ and $ {\rho}(x)$ should be specified.
To avoid our derivation becoming unnecessarily too lengthy,
    we suppose that $ {T}(x)$ and $ {\rho}(x)$
    have the same formal dependence, meaning that ${\cal G}$ in Equation~(\ref{eq_profile_T_overall})
    takes the same form as ${\cal F}$ in Equation~(\ref{eq_profile_rho_overall}).
In addition, we will adopt a number of choices for ${\cal F}$,
\begin{eqnarray}
\label{eq_TL_profile}
   {\cal F}(\varepsilon_{\rm i}, \varepsilon_{\rm e}; x) =
   \left\{
   \begin{array}{ll}
   \varepsilon_{\rm i}-\displaystyle\frac{\varepsilon_{\rm i}-\varepsilon_{\rm e}}{l}\left(x-R+\displaystyle\frac{l}{2}\right),   & {\rm linear},
   \\[0.3cm]
   \varepsilon_{\rm i}-\displaystyle\frac{\varepsilon_{\rm i}-\varepsilon_{\rm e}}{l^2}\left(x-R+\displaystyle\frac{l}{2}\right)^2,& {\rm parabolic},\\[0.3cm]
   \varepsilon_{\rm e}-\displaystyle\frac{\varepsilon_{\rm e}-\varepsilon_{\rm i}}{l^2}\left(x-R-\displaystyle\frac{l}{2}\right)^2,    & {\rm inverse-parabolic} .
   \end{array}
   \right.
\end{eqnarray}
Figure~\ref{fig_profile} uses the transverse density
    distribution as an example to show
    the different choices for ${\cal F}(\varepsilon_{\rm i}, \varepsilon_{\rm e}; x)$,
    where we arbitrarily choose $\rho_{\rm i}/\rho_{\rm e} =50$ and $l/R=1$.

With ${\cal F}$ at hand, we can readily evaluate the coefficients $C_n$ and $V_n$ with Equations~(\ref{eq_coef_Cn})
    and (\ref{eq_coef_Vn}).
Then with the aid of Appendix \ref{sec_app_coef}, the coefficients $a_n$ and $b_n$ can be readily evaluated,
    making it then possible to evaluate the left-hand side (LHS) of the DR~(\ref{eq_DR}).
Restricting us to the applications to standing fast modes in the corona,
    we solve the DR (\ref{eq_DR})
    for complex-valued angular frequencies $\omega$ at a given real-valued longitudinal wavenumber $k$.
Similar to Paper II, we truncate the infinite series expansion
    in Equation (\ref{eq_def_y1y2_expansion}) up to $N=101$.
To speed up the numerical computations, we also reformulate the coefficients in Appendix \ref{sec_app_coef_general}
    such that only
    2-fold summations need to be evaluated
    (see Appendix~\ref{sec_app_coef_para} for details).
We have made sure that adopting an even larger $N$ does not introduce any discernible difference.
The end result is that, once an ${\cal F}(\varepsilon_{\rm i}, \varepsilon_{\rm e}; x)$ is chosen,
    the computations will yield a dimensionless angular frequency $\omega R/v_{\rm Ai}$ that can be formally expressed as
\begin{eqnarray}
    \frac{\omega R}{v_{\rm Ai}} = {\cal H}\left(kR,~\frac{l}{R}, ~\frac{\rho_{\rm i}}{\rho_{\rm e}},~\beta_{\rm i},~\beta_{\rm e}\right) 
\label{eq_omega_formal}
\end{eqnarray}
    for some function ${\cal H}$,
    where $\beta_{\rm i,e} = 2 c_{\rm si,e}^2/(\gamma v_{\rm Ai,e}^2)$.
Throughout this manuscript, the plasma beta in the external medium is fixed at
   $\beta_{\rm e}=0.01$, which is reasonable for a typical coronal environment.
In addition, we will examine only fast standing modes in typical coronal structures, by which
   we mean that the density contrast $\rho_{\rm i}/\rho_{\rm e}$ ranges from $2$ to $200$,
   and the length-to-half-width-ratio $L/R$ ranges between $5$ and $100$.
Note that these values are typical of flare loops and active region loops \citep[e.g.,][]{2004ApJ...600..458A}.
Note further that the longitudinal wavenumber $k$ is related to the structure length
   $L$ by $k = \pi/L$ because we
   shall consider only the longitudinal fundamental mode
   {since it is the mode that is most commonly observed.}

Before proceeding, let us remark that the process of deriving the DR may have introduced some spurious solutions.
As suggested by e.g., \citet{2007SoPh..246..231T} who examined fast modes in cold tubes,
   whether the solutions are physically relevant can be established by performing
   time-dependent simulations on fast modes to address whether the solutions
   play a role in determining the temporal evolution of the system.
Similar to paper II, we have also employed PLUTO, a modular MHD code~\citep{2007ApJS..170..228M},
   to examine both kink and sausage modes in magnetic slabs for which the equilibrium parameters
   are specified by Equation~(\ref{eq_TL_profile}).
This extensive validation study, not shown here,
   indicates that all the solutions to be presented
   are physically relevant.
In addition, the values for the frequencies of fast modes derived from these time-dependent simulations
   are all in close agreement with what we find with the eigen-mode analysis.

\subsection{Effects of a finite beta}
\label{sec_sub_beta_effects}

\subsubsection{Overview of Dispersion Diagrams}
\label{sec_sub_sub_DR}
\begin{figure}
\centering
 \includegraphics[width=0.9\columnwidth]{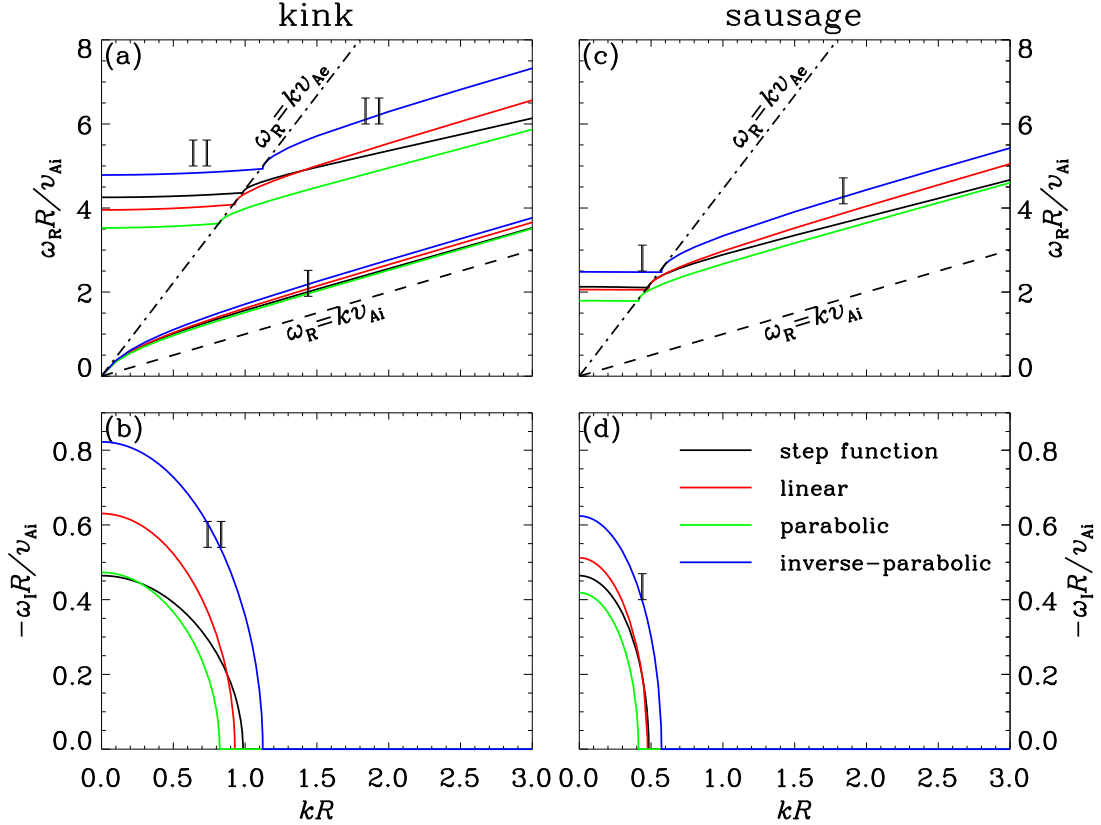}
 \caption{
 Dispersion diagrams for fast kink (left) and sausage
    (right) modes in nonuniform coronal slabs with a finite beta.
 The real ($\omega_{\rm R}$, the upper row) and imaginary
    ($\omega_{\rm I}$, lower row) parts of the angular frequency are shown as functions
    of the real-valued longitudinal wavenumber $k$.
 These solutions are found by solving the dispersion relation
    (Equation (\ref{eq_DR})) for three different profiles, represented by the curves in different colors.
 The corresponding result for a step-function profile is given
    by the black solid curves for comparison.
 In (a) and (c), the black dash-dotted (dashed) lines represent
    $\omega_{\rm R}=kv_{\rm Ae}$ ($\omega_{\rm R}=kv_{\rm Ai}$).
 The former separates the trapped (to its right) from
    leaky (left) regimes, and the latter is the lower limit of $\omega_{\rm R}$ in coronal slabs.
 Here the width of the transition layer $l=R$,
    the density contrast $\rho_{\rm i}/\rho_{\rm e}=10$ and the internal beta $\beta_{\rm i}=1$.
 The external plasma beta is fixed at 0.01.
}
 \label{fig_DR}
\end{figure}

To lay the context for future examinations, this section provides an overview
    of the dispersion diagrams for various profiles in the TL.
Figure \ref{fig_DR} shows the dependence of the real ($\omega_{\rm R}$,
    the upper row) and imaginary ($\omega_{\rm I}$, lower) parts of
    the angular frequency on the longitudinal wavenumber $k$
    for kink (the left column) and sausage (right) modes.
Here we choose $[l/R,~\rho_{\rm i}/\rho_{\rm e},~\beta_{\rm i}]=[1.0, 10, 1.0]$.
Note that $-\omega_{\rm I}$ is plotted instead of $\omega_{\rm I}$
    because $\omega_{\rm I}\leq 0$.
The results for different profiles are
    displayed in different colors as labeled in Figure \ref{fig_DR}(d).
For comparison, the black solid curves represent
    the corresponding results for the top-hat profile ($l/R=0$).
In Figures \ref{fig_DR}a and \ref{fig_DR}c, the dash-dotted and dashed
    lines represent $\omega_{\rm R}=kv_{\rm Ae}$ and $\omega_{\rm R}=kv_{\rm Ai}$, respectively.
The former separates the trapped (to its right) from the
    leaky (left) regime, and the latter is the lower limit
    that $\omega_{\rm R}$ attains for fast modes.
For the kink modes, we examine only the first two branches (labeled I and II
    in the left column), and call them ``kink I'' and ``kink II'' for brevity.
Likewise, we examine only the lowest-order sausage modes, to be called ``sausage I'' from here onwards.

Some general features are evident from Figure~\ref{fig_DR}.
First, for kink and sausage modes alike, the real parts of their angular frequencies
    always lie above the dashed lines, meaning that these modes are always body modes.
In fact, this can be said for all fast modes {when $v_{\rm Ae} > v_{\rm Ai}$}, 
    regardless of their transverse harmonic number even though
    only the first several branches are shown here.
Second, with the exception of kink I, all modes become leaky when $k$ is below some critical value
    as evidenced by the non-zero values of $\omega_{\rm I}$ in the lower panels.
This is well-known for top-hat profiles~\citep[e.g.,][]{1982SoPh...76..239E}, and the numerical results
    shown here demonstrate that this holds for rather arbitrary choices of
    temperature and density profiles in the TL.
Furthermore, the same behavior holds for rather arbitrary choices of the plasma beta
    and was seen in paper I where $\beta$ was taken to be zero.

Despite the similarities in the overall behavior to the top-hat case,
    choosing different profiles in the TL can considerably impact
    the specific values of the angular frequency.
Take kink II in the limit $kR \rightarrow 0$ for instance.
While $[\omega_{\rm R}, -\omega_{\rm I}]R/v_{\rm Ai}$ reads $[4.25, 0.464]$
    for top-hat profiles (the black solid curves in Figs.~\ref{fig_DR}a and \ref{fig_DR}b),
    it attains $[3.53, 0.473]$ ($[4.79, 0.822]$) for
    the profile labeled ``parabolic'' (``inverse-parabolic''), given
    by the green (blue) curves.
This happens in conjunction with the rather sensitive profile dependence of
    the critical wavenumber $k_{\rm c}R$.
For the top-hat, parabolic and inverse-parabolic profiles,
    $k_{\rm c}R$ reads $0.99$, $0.83$, and $1.13$, respectively.
From an observational perspective, this means that
    the largely unknown transverse distribution across coronal structures
    can in principle be inferred by measuring such parameters as
    the periods and damping times of standing fast modes.
In practice, this approach has been extensively employed,
    albeit almost exclusively based on the theoretical results
    found in cold MHD \citep[e.g.,][]{2007A&A...463..333A,
    2008A&A...484..851G,2014ApJ...781..111S,2016SoPh..291..877G}.
Therefore what remains to be examined is how the dispersive properties
    of fast oscillations depend on plasma beta for a chosen profile.
In what follows, we will examine how the periods ($P=2\pi/\omega_{\rm R}$),
    damping times ($\tau = 1/|\omega_{\rm I}|$) and the critical wavenumbers ($k_{\rm c}R$)
    depend on plasma beta for the profiles we chose.
Fortunately, we will need only to show the results for an arbitrarily chosen profile
    (the parabolic one, to be specific), because the overall beta dependence remains the same
    for the rest of {the} profile choices.

\subsubsection{First Branch of Fast Kink Modes}
\begin{figure}
\centering
 \includegraphics[width=0.9\columnwidth]{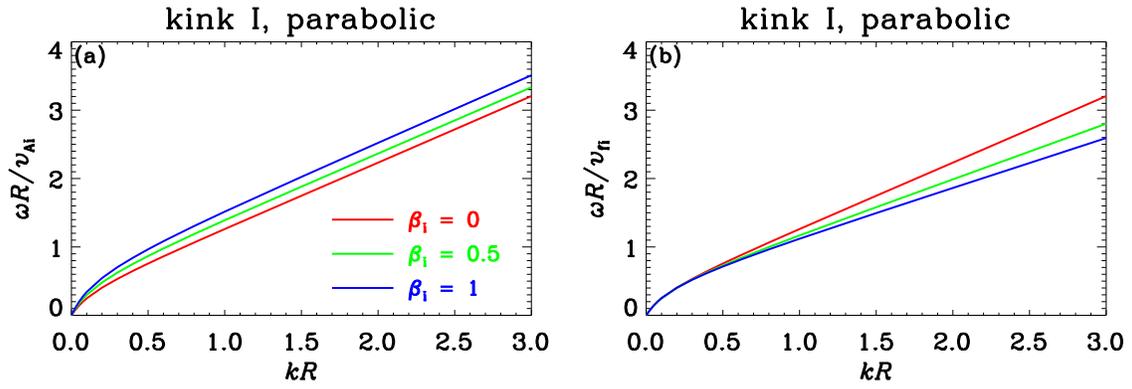}
 \caption{
Dispersion diagrams for the first branch of standing kink
    modes in nonuniform coronal slabs with a finite beta.
The angular frequency $\omega$ in units of $v_{\rm Ai}/R$ (panel a)
    and $v_{\rm fi}/R$ (panel b) are shown as functions of the longitudinal wavenumber $k$.
The red, green, and blue curves correspond to
    $\beta_{\rm i}=0$, $0.5$, and $1$, respectively.
Here the transverse profile is
    {parabolic, the width of the transition layer is $l=R$,
    and the density contrast is $\rho_{\rm i}/\rho_{\rm e}=10$}.
}
 \label{fig_DR_firstkink}
\end{figure}

This is the simplest to examine given that kink I is always trapped ($\omega$ is always real).
Figure~\ref{fig_DR_firstkink} presents how $\omega$ depends
    on the longitudinal wavenumber $kR$ by examining $\omega$ measured in units of both $v_{\rm Ai}/R$ (the left panel)
    and $v_{\rm fi}/R$ (right).
Here $[l/R,~\rho_{\rm i}/\rho_{\rm e}]$ is fixed at $[1.0, 10]$,
    but a number of different values for the internal plasma beta $\beta_{\rm i}$ are examined
    (see curves in different colors).
From Figure~\ref{fig_DR_firstkink} one sees that regardless of $\beta_{\rm i}$, the angular frequency
    $\omega$ always increases with $kR$, be it normalized by the transverse fast or Alfv\'en time.
However, Figure~\ref{fig_DR_firstkink}b indicates that the curves show a considerable weaker dependence on $\beta_{\rm i}$
    than in Figure~\ref{fig_DR_firstkink}a.
In fact, it is hard to tell the curves apart when $k R \lesssim 0.6$.
This means that if we reformulate Equation (\ref{eq_omega_formal}) to
\begin{eqnarray}
    \frac{\omega R}{v_{\rm fi}} = {\cal L}\left(kR,~\frac{l}{R}, ~\frac{\rho_{\rm i}}{\rho_{\rm e}},~\beta_{\rm i},~\beta_{\rm e}\right)~,
\label{eq_omega_formal_fi}
\end{eqnarray}
    then the function ${\cal L}$ possesses a much weaker $\beta_{\rm i}$ dependence
    than ${\cal H}$, as long as the length-to-half-width-ratio $L/R \gtrsim \pi/0.6 \approx 5$.
In some sense this is not surprising because Equation~(\ref{eq_omega0}) suggests that
    for fast modes in infinitely thin structures,
    $\omega R/v_{\rm fi}$ does not depend on $\beta_{\rm i}$ at all for top-hat profiles.
What Figure~\ref{fig_DR_firstkink} indicates is that the dependence on $\beta_{\rm i}$
    remains weak for continuous transverse profiles provided that the coronal structures
    are not unrealistically thick.

Does this conclusion hold for other choices of the density contrast?
We examine this by asking how much the period $P$
    may differ from its cold MHD counterpart ($\beta_{\rm i}= \beta_{\rm e}=0$).
To be more specific, Equation~(\ref{eq_omega_formal_fi}) suggests that
    $P$  in units of
    $R/v_{\rm fi}$ at a given pair of $[\rho_{\rm i}/\rho_{\rm e},~L/R]$
    is a function of $\beta_{\rm i}$ only when $l/R$ is fixed.
Let this value be denoted by $P^{\beta\neq0}$.
At the same given $[\rho_{\rm i}/\rho_{\rm e},~L/R]$, we then
    evaluate $P$ in units of $R/v_{\rm fi}$ in the cold MHD limit
    by solving the corresponding DR (Equation 17 in Paper I).
Let $P^{\rm cold}$ denote this value.
We now define $\delta P$ to be the maximal relative
    difference between the finite-$\beta$
    and cold MHD results when $\beta_{\rm i}$ varies between $0$ and~$1$.
In other words,
 \begin{equation}\label{eq_delta_P}
  \delta P\equiv \max\left|\displaystyle\frac{P^{\beta\neq0}(\beta_{\rm i}\in[0,~1])}{P^{\rm cold}}-1\right|~,
     \end{equation}
    which now depends on $\rho_{\rm i}/\rho_{\rm e}$ and $L/R$ only.

\begin{figure}
\centering
 \includegraphics[width=0.8\columnwidth]{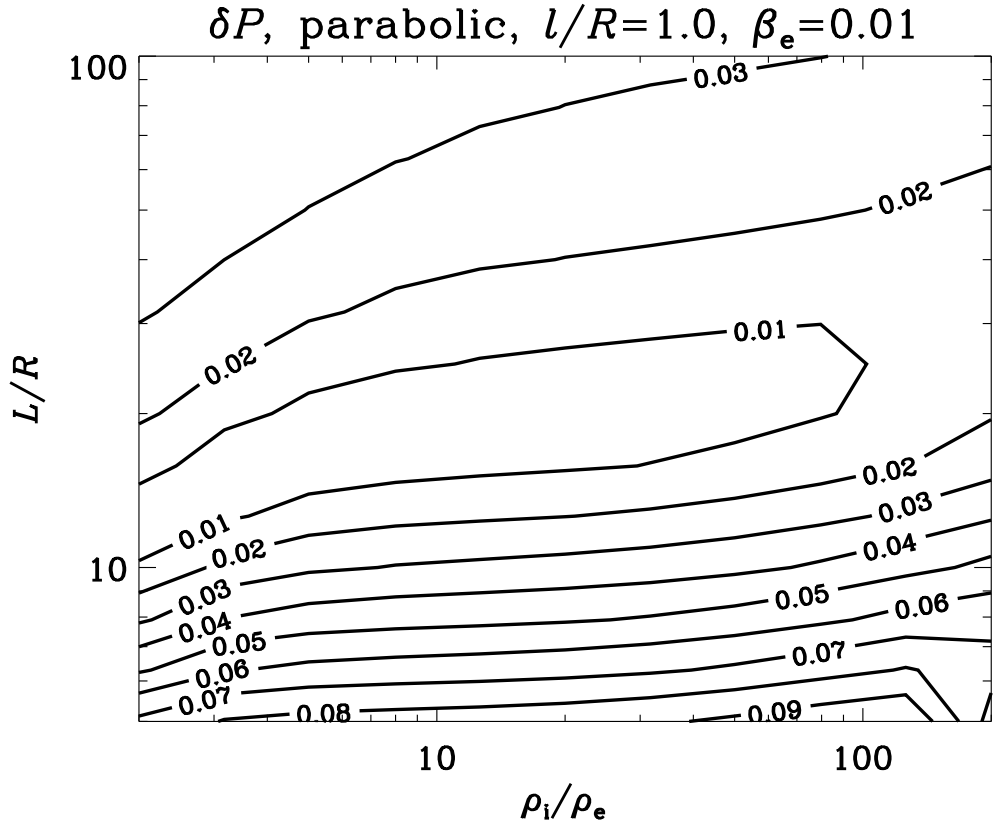}
 \caption{
 Distributions of $\delta P$ for the first branch of kink modes
    in the space spanned by
    $\rho_{\rm i}/\rho_{\rm e}$ and $L/R$ for parabolic profiles with $l/R=1$.
 Here $\delta P$ evaluates the maximal difference of the
    period $P$ relative to the cold MHD result at a given pair $[\rho_{\rm i}/\rho_{\rm e},~L/R]$ when $\beta_{\rm i}$ varies between 0 and 1.
 See text for details.
}
 \label{fig_contour_firstkink}
\end{figure}

Figure \ref{fig_contour_firstkink} shows how $\delta P$
    is distributed in the $[\rho_{\rm i}/\rho_{\rm e},~L/R]$ space for $l/R = 1$.
One sees that $\delta P$ is consistently less than $10\%$ despite the considerable variations in
    both $\rho_{\rm i}/\rho_{\rm e}$ and $L/R$.
Note that our cold MHD results have also shown that the periods $P$ for fast modes pertaining to kink I
    do not differ much if we change one TL profile to another (paper I).
In fact, for combinations of $[\rho_{\rm i}/\rho_{\rm e}$, $L/R]$ in the same range as examined here,
    Figure~3 in paper I demonstrates that $P$ is rather insensitive to the dimensionless layer width $l/R$.
It then follows that as far as kink I is concerned, one may adopt the top-hat results
    as a reasonable starting point when fast modes pertaining to kink I are put to seismological use.
This is good news for solar MHD seismology because the top-hat results are much less complicated (compare Eq.~\ref{eq_DR} with \ref{eq_DR_tophat}).
On top of that, the detailed form of the transverse distributions of the physical parameters
    proves difficult to infer and suffers from considerable uncertainties~\citep[e.g.,][]{2014A&A...565A..78A,2015ApJ...811..104A, 2017A&A...600L...7P}.

\subsubsection{Second Branch of Kink Modes}

\begin{figure}
\centering
 \includegraphics[width=0.8\columnwidth]{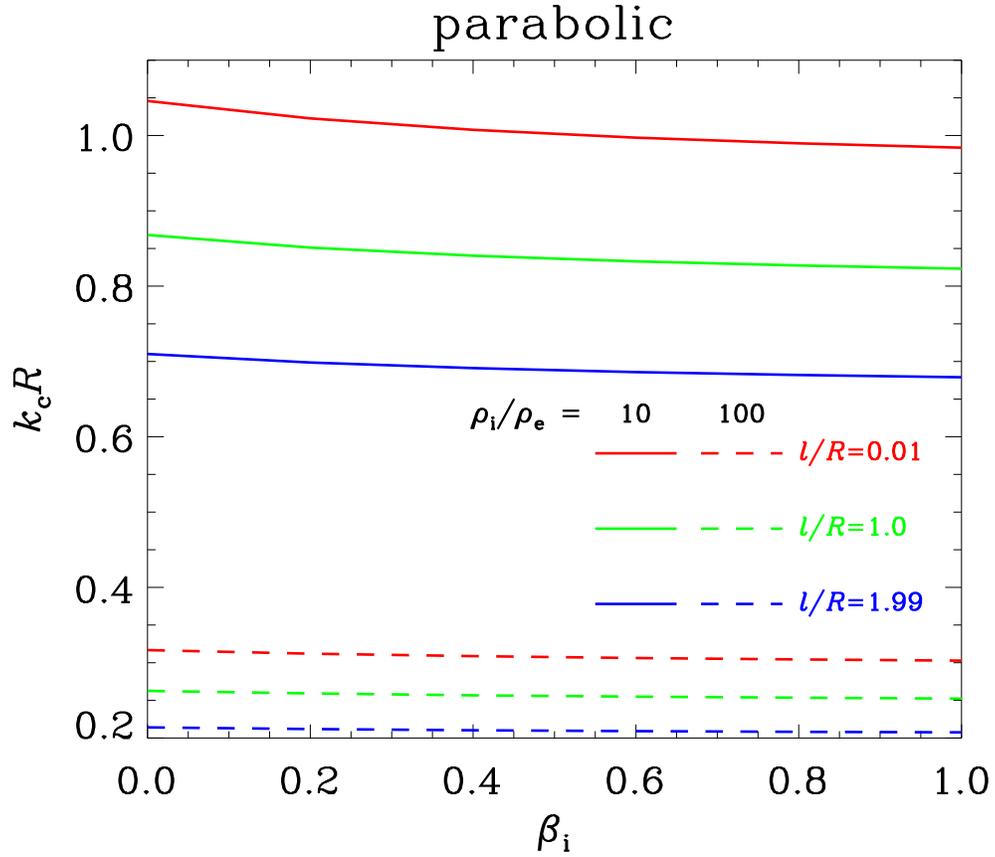}
 \caption{
 Dependence on the internal plasma beta ($\beta_{\rm i}$)
    of critical wavenumbers $k_{\rm c}$ of the second branch of kink modes in transversely continuous slabs with parabolic profiles.
 A number of combinations for the density contrast
    $\rho_{\rm i}/\rho_{\rm e}$ and 
    {dimensionless layer width} $l/R$ are examined as labeled.
 }
 \label{fig_kc_secondkink}
\end{figure}

The examination of the beta dependence of the dispersive properties of kink II
    is substantially more complicated because now the modes can become leaky for sufficiently
    {thin structures}.
Figure~\ref{fig_kc_secondkink} examines the influence of a finite $\beta_{\rm i}$
    on $k_{\rm c}$ for parabolic profiles.
A number of combinations $[l/R,~\rho_{\rm i}/\rho_{\rm e}]$
    are examined and given by the different colors and line styles.
It is clear that the most important factor that
    influences $k_{\rm c} R$ is the density contrast $\rho_{\rm i}/\rho_{\rm e}$.
Take the green curves pertaining to $l/R=1$ for instance.
One sees that for all the $\beta_{\rm i}$ values examined, the critical wavenumber $k_{\rm c}R$
    substantially decreases when $\rho_{\rm i}/\rho_{\rm e}$ increases
    from $10$ (the solid curve) to $100$ (dashed).
On the other hand, examining any individual curve indicates that
    $k_{\rm c}R$ is not sensitive to $\beta_{\rm i}$,
    which is particularly true for large density contrasts.
Even for the smaller $\rho_{\rm i}/\rho_{\rm e}=10$ (the solid curves),
    increasing $\beta_{\rm i}$ leads to a decrease in $k_{\rm c}$
    by no more than $5.93\%$ for all the layer widths considered.
This insensitivity to $\beta_{\rm i}$ of $k_{\rm c}R$
    can be partly understood from Equation (\ref{eq_kc_tophat}), which pertains to top-hat profiles.
Reformulating it to make the $\beta_{\rm i}$ dependence more apparent, one finds that
\begin{eqnarray}
     k_{\rm c} R = g_2 \sqrt{-1+\frac{\rho_{\rm ie}^2 (1+\beta_{\rm i})^2}
         {[\rho_{\rm ie}(1+\beta_{\rm i})-(1+\beta_{\rm e})][\rho_{\rm ie}(1+\beta_{\rm i})-(\gamma\beta_{\rm i}/2)(1+\beta_{\rm e})]}}
     \label{eq_kc_tophat_dmless}
\end{eqnarray}
    with $g_2 = \pi$ (see Eq.~\ref{eq_gl}).
Here we also used the shorthand notation $\rho_{\rm ie}=\rho_{\rm i}/\rho_{\rm e}$.
Given that $\beta_{\rm e} \ll 1$ for a typical coronal ambient,
    $k_{\rm c}$ can be approximated to within $\sim 10\%$ by
\begin{eqnarray}
     k_{\rm c} R \approx g_2 \sqrt{\frac{1+\gamma\beta_{\rm i}/2}{\rho_{\rm ie}(1+\beta_{\rm i})}} ,
     \label{eq_kc_apprx_tophat}
\end{eqnarray}
     when $\rho_{\rm ie}\gtrsim 5$.
Equation (\ref{eq_kc_apprx_tophat}) suggests that the reason for the insensitive $\beta_{\rm i}$ dependence of $k_{\rm c}R$
     is twofold.
One is the appearance of the square root,
     and the other is that $\gamma/2$ is close to unity.
What Figure~\ref{fig_kc_secondkink} suggests is that this weak dependence on $\beta_{\rm i}$ persists
     for continuous profiles.

\begin{figure}
\centering
 \includegraphics[width=0.7\columnwidth]{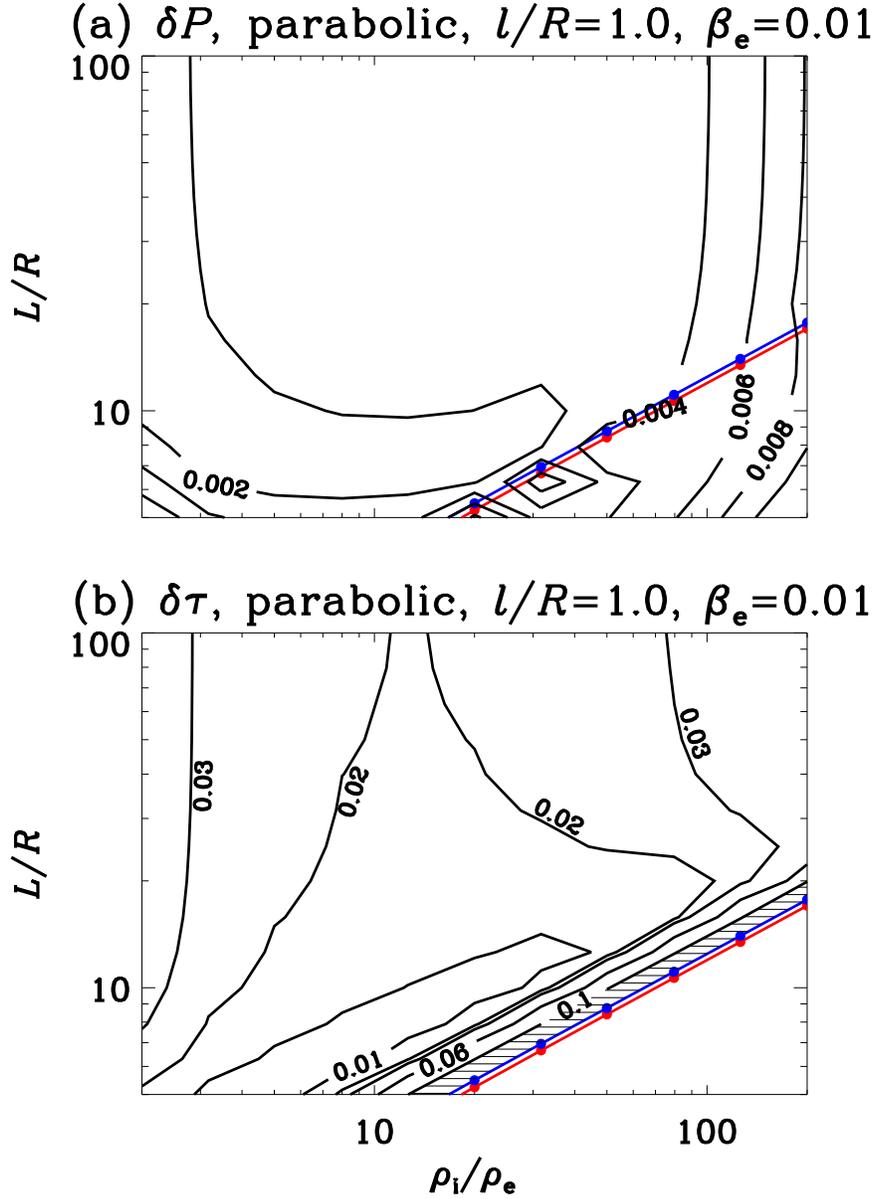}
 \caption{
 Distributions of $\delta P$ and $\delta\tau$ for the second branch of kink modes
    in the space
    spanned by $\rho_{\rm i}/\rho_{\rm e}$ and $L/R$ for parabolic profiles with $l/R=1$.
 Here $\delta P$ ($\delta \tau$) evaluates the maximal difference of the
    period $P$ (damping time $\tau$) relative to the cold MHD result at a given pair $[\rho_{\rm i}/\rho_{\rm e},~L/R]$ when $\beta_{\rm i}$ varies between 0 and 1.
 In addition, the red and blue lines represent the lower
    and upper limits of the critical length-to-half-width-ratios at a given density contrast.
The hatched area in panel (b) represents where $\delta\tau$
    exceeds $10\%$.
See text for details.
}
 \label{fig_contour_secondkink}
\end{figure}

How about the $\beta_{\rm i}$ dependence of the angular frequencies of kink II?
We tackle this issue with the same approach as for kink I.
It is just that now in addition to the periods ($P$), the damping times ($\tau$) are also relevant
     {because these modes can be in the leaky regime}.
For this purpose, we also define $\delta \tau$ as
\begin{equation}\label{eq_delta_tau}
      \delta \tau\equiv \max\left|\displaystyle\frac{\tau^{\beta\neq0}(\beta_{\rm i}\in[0,~1])}{\tau^{\rm cold}}-1\right|~,
\end{equation}
    which is identical in form to Equation~(\ref{eq_delta_P}) except that $P$ is replaced with $\tau$.
Figure~\ref{fig_contour_secondkink} then presents the
    distributions in the $[\rho_{\rm i}/\rho_{\rm e},~L/R]$ space of
    both $\delta P$ (panel a) and $\delta\tau$ (panel b)
    for a dimensionless layer width $l/R = 1$.
The red and blue curves represent the lower and
    upper limits of $(L/R)_{\rm c}$ when $\beta_{\rm i}$ varies from $0$ to $1$,
    {with $(L/R)_{\rm c} = \pi/(k_{\rm c}R)$ denoting the critical length-to-half-width-ratio.}
Trapped (leaky) modes lie on the right (left) of these lines,
    and hence $\delta\tau$ is undefined in the lower-right corner
    of Figure~\ref{fig_contour_secondkink}b.
One sees that the red and blue curves differ little, which is not surprising given
    the insensitive dependence of $k_{\rm c}R$ on $\beta_{\rm i}$.

What is more interesting is that neither $P$ nor $\tau$ is sensitive to $\beta_{\rm i}$
    as long as they are measured in units of the transverse fast time $R/v_{\rm fi}$.
Figure~\ref{fig_contour_secondkink}b indicates that $\delta \tau$ exceeds $10\%$ only in the
    immediate vicinity of the red or blue curve, as represented by the hatched portion.
Regarding the periods, Figure~\ref{fig_contour_secondkink}a indicates that $\delta P$ is consistently smaller
    than $1.2\%$.
This $\beta_{\rm i}$ dependence is even weaker than for kink I (see Figure~\ref{fig_contour_firstkink}).
However, it should be noted that in cold MHD the specific values of $P$ and $\tau$
    are rather sensitive to the parameters characterizing the TL, i.e.,
    $\rho_{\rm i}/\rho_{\rm e}$ and $l/R$.
At a given pair of $\rho_{\rm i}/\rho_{\rm e}$ and $l/R$,
    they are also considerably influenced by how the TL profile is prescribed (paper I).
Therefore what Figure~\ref{fig_contour_secondkink} indicates is that, if $P$ and $\tau$ of kink II
    are to be used for seismological purposes, then one can use the much simpler cold MHD theory
    as presented in paper I.
The difference between kink I and kink II is that, while one can use the even simpler theory for top-hat profiles to evaluate $P$
    for kink I, the details of the TL profiles have to be taken into account when $P$ and $\tau$ are
    evaluated for kink II.

\subsubsection{First Branch of Sausage Modes}
\begin{figure}
\centering
 \includegraphics[width=0.9\columnwidth]{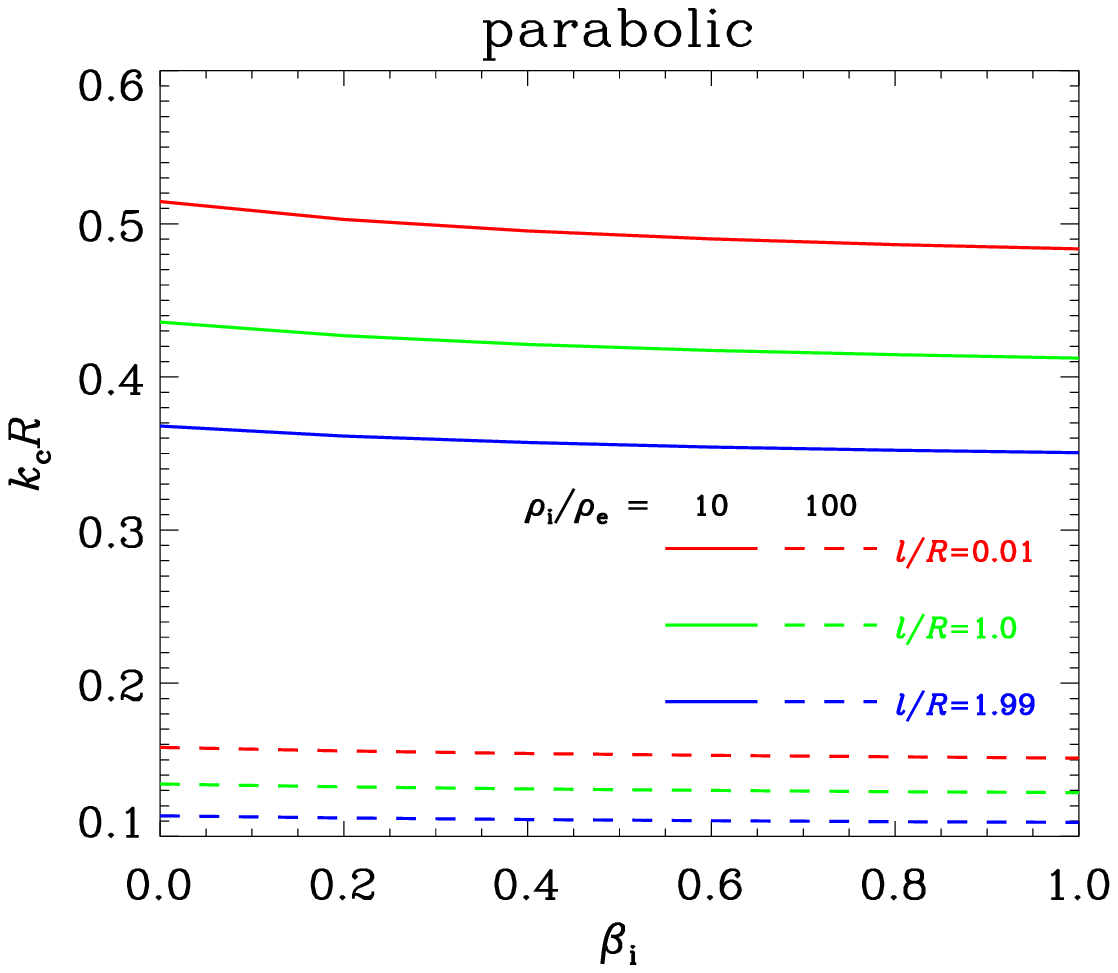}
 \caption{
 Similar to Figure \ref{fig_kc_secondkink} but for the first branch of sausage modes.
}
 \label{fig_kc_firstsausage}
\end{figure}

The influence of $\beta_{\rm i}$ on sausage I can be examined in a manner
    identical to kink II.
Figure~\ref{fig_kc_firstsausage} shows the critical wavenumber
    $k_{\rm c} R$ as a function of $\beta_{\rm i}$ for parabolic profiles
    with a number of different combinations of $\rho_{\rm i}/\rho_{\rm e}$ and $l/R$.
Evidently, the behavior of $k_{\rm c} R$ is qualitatively the same as for kink II.
Once again, by far the most important factor that influences $k_{\rm c}R$
    is the density contrast,
    and the role played by $\beta_{\rm i}$ is marginal to say the most.
Similar to kink II, this insensitive dependence on $\beta_{\rm i}$ can be partly understood from
    Equation~(\ref{eq_kc_apprx_tophat}) pertinent to top-hat profiles.
The only difference is that $g_2$ needs to be replaced with $\pi/2$  (see Eq.~\ref{eq_gl}).
Note that for top-hat profiles, this weak dependence of $k_{\rm c}R$ on $\beta_{\rm i}$ was already shown for
    sausage modes by~\citet{2009A&A...503..569I}.
What Figure~\ref{fig_kc_firstsausage} shows is that
    this weak dependence persists for continuous profiles as well.

\begin{figure}
\centering
 \includegraphics[width=0.7\columnwidth]{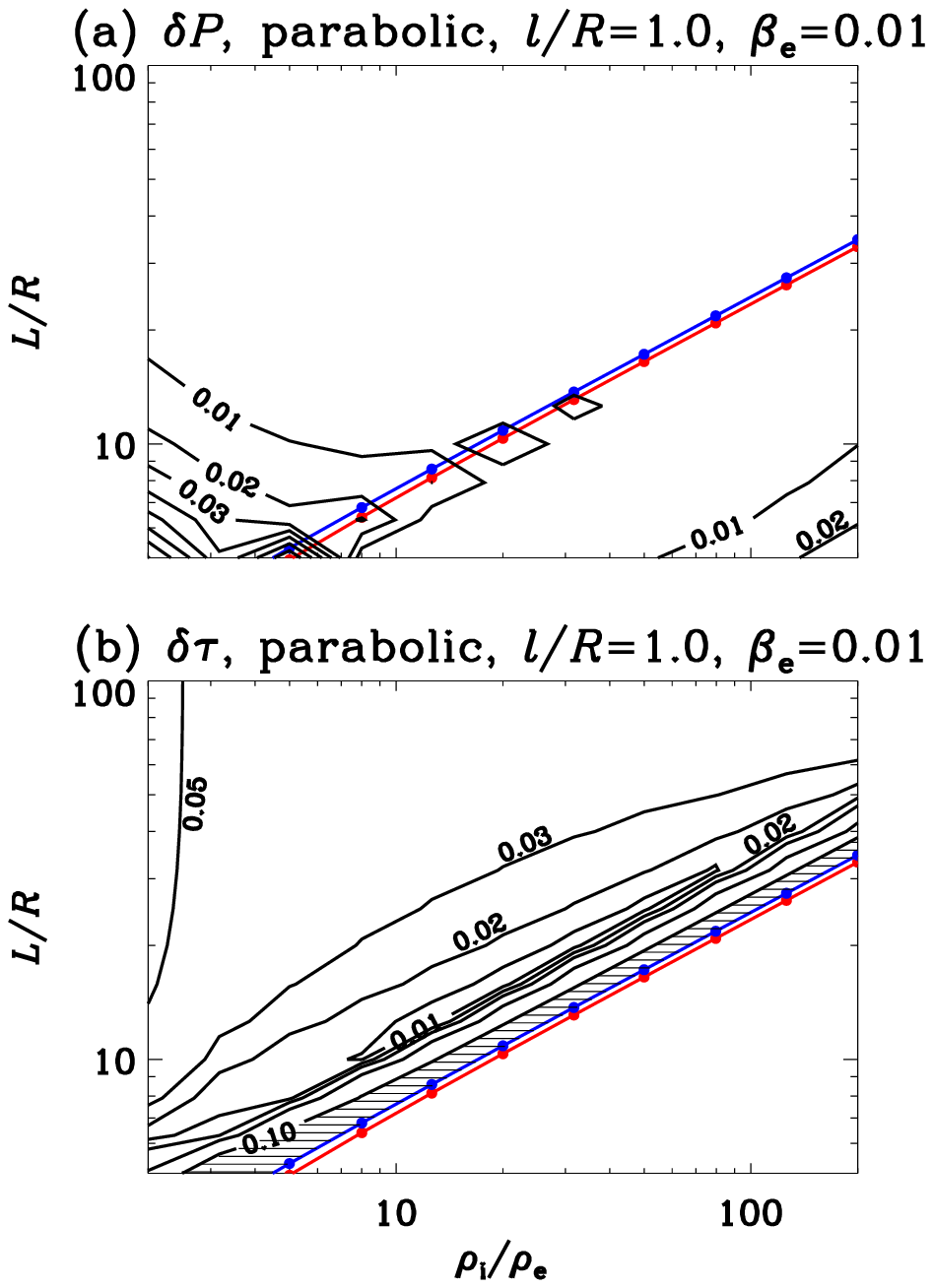}
 \caption{
Similar to Figure \ref{fig_contour_secondkink} but for the first branch of sausage modes.
}
 \label{fig_contour_firstsausage}
\end{figure}

The $\beta_{\rm i}$ dependence of the periods $P$ and damping times $\tau$
    is brought out also by examining how they differ from the cold MHD results.
Similar to Figure~\ref{fig_contour_secondkink}, now $\delta P$ and $\delta \tau$
    are presented in Figure~\ref{fig_contour_firstsausage} as functions of $\rho_{\rm i}/\rho_{\rm e}$ and $L/R$.
Comparing Figure~\ref{fig_contour_firstsausage}b with Figure~\ref{fig_contour_secondkink}b indicates
    that the hatched portion where $\delta \tau$ exceeds $10\%$
    is somehow broader than for kink II.
Nonetheless, in the majority of the parameter space, the influence of $\beta_{\rm i}$ on $\tau$ tends to be marginal to say the most.
This weak $\beta_{\rm i}$ dependence is even more pronounced for the periods, for which
    Figure~\ref{fig_contour_firstsausage}a indicates that $P$ differs from its cold MHD counterpart by
    no more than $3\%$.
Now recall that our paper I has demonstrated that in cold MHD, the periods and damping times for sausage I
    are sensitive to the profile prescriptions.
Therefore our conclusion regarding sausage I is identical to kink II, namely
    the corresponding $P$ and $\tau$ can be evaluated with the DR in cold MHD
    (Equation~17 in paper I) by properly accounting for the transverse density distribution.
The effect of a finite $\beta_{\rm i}$ is secondary, as long as $P$ and $\tau$ are measured
    in units of the transverse fast time.

\section{SUMMARY AND CONCLUDING REMARKS}
\label{sec_conclusion}
This study has been motivated by the apparent lack of a theoretical examination on
    the combined effects on fast standing modes of a finite plasma beta inside
    and a continuous distribution of equilibrium parameters across solar coronal slabs.
In the framework of ideal MHD with finite gas pressure, we worked out a rather generic
    dispersion relation (Equation~\ref{eq_DR}) governing fast modes
    in magnetic slabs for which the transverse profiles comprise a uniform core,
    a uniform external medium, and a transition layer (TL) sandwiched in between.
The profiles in the TL are allowed to be rather arbitrary.
We have restricted our attention to the first several branches of fast modes, which are of
    observational interest in most cases.
The influence of a finite plasma beta on the dispersive properties of these modes is brought out
    by examining how the periods ($P$), damping times ($\tau$), and critical longitudinal wavenumbers ($k_{\rm c}$)
    are affected.
Our numerical results indicated that for parameters typical of coronal structures,
    the influence due to a finite plasma beta is at most marginal,
    as long as both $P$ and $\tau$ are measured in units of the transverse fast time.
Putting these results together with our cold MHD results as presented in \citet{2015ApJ...814...60Y},
    we conclude that for the first branch of sausage modes and second branch of kink modes alike,
    $P$ and $\tau$ can be evaluated with the theoretical results found for cold slabs provided that
    the transverse profiles are properly accounted for and the transverse Alfv\'en time in cold MHD
    is replaced with the transverse fast time.
For the first branch of kink modes, one can use the even simpler theory for cold slabs with top-hat density profiles
    because their periods are not sensitive to either the layer width
    or how the density profile is prescribed in the TL.

This study can be extended in a number of ways.
For instance, by accounting for the out-of-plane propagation, the interesting physics of resonant coupling of
    fast kink waves to the Alfv\'en continuum can be examined~\citep[e.g.,][]{2007SoPh..246..213A}.
In this case, a singular series expansion, 
    rather than the regular series expansion that was adopted here, is necessary to solve the governing differential equation
    (see~\citeauthor{2013ApJ...777..158S}~\citeyear{2013ApJ...777..158S} for the application of this approach in the cylindrical geometry).
When this approach is implemented, the resonant coupling of slow modes to the cusp continuum can also be addressed,
    and the finite gas pressure is expected to play an important role.
Furthermore, by allowing the physical parameters in the environment to be asymmetric about the slab axis,
    one will be able to examine how a continuous transverse profile affects the waves modes, which
    can no longer be strictly classified into kink and sausage modes~\citep{2017SoPh..292...35A}.
This is expected to find applications to structures in the lower solar atmosphere, those close to the magnetic canopy
    for instance.

\acknowledgments
{We thank the referee for his/her constructive comments.}
This work is supported by
    the National Natural Science Foundation of China (BL:41674172, 11761141002, and 41474149, SXC:41604145, HY:41704165),
    and by the Provincial Natural Science Foundation of Shandong via Grants JQ201212 (BL) and ZR2016DP03 (HY).

\bibliographystyle{apj}
\bibliography{slab_finite_beta}

     % Checking: look if the file containing the ``\bibitem'' exits
     %           so check if the .bbl file exist (bibTeX compilation)
\IfFileExists{\jobname.bbl}{} {\typeout{}
\typeout{****************************************************}
\typeout{****************************************************}
\typeout{** Please run "bibtex \jobname" to obtain} \typeout{**
the bibliography and then re-run LaTeX} \typeout{** twice to fix
the references !}
\typeout{****************************************************}
\typeout{****************************************************}
\typeout{}}

%%%%%%%%%%%%%%%%%%%%%%%%%%%%%%%%%%%%%%%%%%%

\begin{center}
{\bf APPENDIX}
\end{center}

\appendix
\section{Coefficients in the expressions for $\tilde\xi_{\rm tr,1}(\epsilon)$ and $\tilde\xi_{\rm tr,2}(\epsilon)$}
\label{sec_app_coef}

\subsection{Coefficients for general profiles in the transition layer}
\label{sec_app_coef_general}
For general density and temperature profiles in the transition layer
   described in Equations (\ref{eq_profile_rho_overall}) and (\ref{eq_profile_T_overall}),
   the coefficients $a_n$
   and $b_n$ in $\tilde\xi_{\rm tr,1}(\epsilon)=\sum\limits^\infty_{n=0}a_n\epsilon^n$ and $\tilde\xi_{\rm tr,2}(\epsilon)=\sum\limits^\infty_{n=0}b_n\epsilon^n$ are given by
\begin{equation}
  \left\{
  \begin{array}{rcl}
  a_0&=&R \\ [0.3cm]
  a_1&=&0
  \end{array}
  \right. {\rm ~~~and~~~}
  \left\{
  \begin{array}{rcl}
  b_0&=&0 \\ [0.3cm]
  b_1&=&1~.
  \end{array}
  \right.
 \end{equation}
From this point onward, let $\chi$ denote either $a$ or $b$, since both obey the same
    recurrence relations.
The coefficients $\chi_i$ for $i \ge 2$ are then given by
\begin{equation}
   \chi_i=-\displaystyle\frac{D(k,\omega^2)}{i(i-1)C(k,\omega^2)}
\label{eq_chi_gt3}
\end{equation}
where
\begin{equation}
\begin{aligned}
   C(k,\omega^2)&=2\omega^4C^2_0+\gamma\omega^4C_0V_0-2k^2\omega^2C^3_0
   -\gamma k^2\omega^2C^2_0V_0+2\omega^4C_0V_0+\gamma\omega^4V^2_0\\
   &-4k^2\omega^2C^2_0V_0-2\gamma k^2\omega^2V^2_0C_0+2k^4C^3_0V_0+\gamma k^4C^2_0V^2_0
\end{aligned}
 \end{equation}
and
\begin{equation}
\begin{array}{rcl}
   &&D(k,\omega^2)=D_1(k,\omega^2)+D_2(k,\omega^2)+D_3(k,\omega^2)\\[0.3cm]
   &&D_1(k,\omega^2)\\
   &&=\omega^4\sum\limits_{m=0}^{i-3}\sum\limits_{j=0}^{i-2-m}(m+2)(m+1)\left( C_{i-2-j-m}+V_{i-2-j-m}\right)(2C_j+\gamma V_j)\chi_{m+2}\\
   &&-k^2\omega^2\sum\limits_{m=0}^{i-3}\sum\limits_{j=0}^{i-2-m}\sum\limits_{l=0}^{i-2-j-m}(m+2)(m+1)(C_{i-2-l-j-m}+2V_{i-2-l-j-m})(2C_j+\gamma V_j)C_l\chi_{m+2}\\
   &&+k^4\sum\limits_{m=0}^{i-3}\sum\limits_{j=0}^{i-2-m}
   \sum\limits_{l=0}^{i-2-j-m}\sum\limits_{s=0}^{i-2-j-l-m}(m+2)(m+1)C_{i-2-j-l-m-s}V_s C_j(2C_l+\gamma V_l)\chi_{m+2}
   \end{array}
\end{equation}
\begin{equation}
\begin{array}{rcl}
&&D_2(k,\omega^2)\\
&&=\omega^4\sum\limits_{m=0}^{i-2}\sum\limits_{j=0}^{i-2-m}(j+1)(2-\gamma)
(C_{i-2-j-m}V_{j+1}-C_{j+1}V_{i-2-j-m})(m+1)\chi_{m+1}\\[0.3cm]
&&+k^2\omega^2\sum\limits_{m=0}^{i-2}\sum\limits_{j=0}^{i-2-m}
\sum\limits_{l=0}^{i-2-j-m}(m+1)(j+1)(4V_lC_{j+1}+\gamma C_lV_{j+1}+2C_lC_{j+1}-4C_lV_{j+1})C_{i-2-j-l-m}\chi_{m+1}\\[0.3cm]
&&+2k^4\sum\limits_{m=0}^{i-2}\sum\limits_{j=0}^{i-2-m}
\sum\limits_{l=0}^{i-2-j-m}\sum\limits_{s=0}^{i-2-j-l-m}
(j+1)(C_sV_{j+1}-V_sC_{j+1})C_{i-2-j-l-m-s}C_l(m+1)\chi_{m+1}
\end{array}
\end{equation}
\begin{equation}
\begin{array}{lll}
&&D_3(k,\omega^2)\\
&&=\omega^6\sum\limits_{m=0}^{i-2}(2C_{i-2-m}+\gamma V_{i-2-m})\chi_m-k^2\omega^4\sum\limits_{m=0}^{i-2}\sum\limits_{j=0}^{i-2-m}
(2C_j+\gamma V_j)(2C_{i-2-j-m}+V_{i-2-j-m})\chi_m\\
&&+k^4\omega^2\sum\limits_{m=0}^{i-2}\sum\limits_{j=0}^{i-2-m}
\sum\limits_{l=0}^{i-2-j-m}C_{i-2-j-l-m}(2C_j+\gamma V_j)(C_l+2 V_l)\chi_m\\
&&-k^6\sum\limits_{m=0}^{i-2}\sum\limits_{j=0}^{i-2-m}
\sum\limits_{l=0}^{i-2-j-m}\sum\limits_{s=0}^{i-2-j-l-m}C_{i-2-j-l-m-s}C_s(2C_j+\gamma V_j)V_l\chi_m.
\end{array}
\end{equation}

\subsection{Simplified Coefficients for Profiles Specified in Equation (\ref{eq_TL_profile})}
\label{sec_app_coef_para}
Given that the coefficients
     $C_i~(i>2)$ are all zero for the equilibrium
     profiles given by Equation (\ref{eq_TL_profile}),
     we can avoid the most time-consuming part when evaluating the coefficients $a_{i}$ and $b_i$
     by simplifying the 4-fold summations.
For $i\ge 6$,
    the terms $D_1$, $D_2$ and $D_3$ in Equation (\ref{eq_chi_gt3}) can be reformulated such that only 2-fold summations are involved.
In other words,
\begin{equation}
\begin{aligned}
%& D_1(k,\omega^2) \\
D_1(k,\omega^2)&=\sum_{m=0}^{i-3}\sum_{j=0}^{i-2-m}(m+2)(m+1)\chi_{m+2}(2C_j+\gamma V_j)\left[\omega^4(C_{i-2-j-m}+V_{i-2-j-m})\right.\\
&~~~~~~~~~~~~~~~~
\left.-k^2\omega^2C_0(C_{i-2-j-m}+2V_{i-2-j-m})+k^4C^2_0V_{i-2-j-m}\right]\\
&+\sum_{m=0}^{i-3}\sum_{j=0}^{i-3-m}(m+2)(m+1)(2C_j+\gamma V_j)\chi_{m+2}\\
&~~~~~~~~~~~~~~~~\left[2k^4C_0C_1V_{i-3-j-m}-k^2\omega^2C_1(C_{i-3-j-m}+2V_{i-3-j-m})\right]\\
&+\sum_{m=0}^{i-4}\sum_{j=0}^{i-4-m}(m+2)(m+1)(2C_j+\gamma V_j)\chi_{m+2}\\
&~~~~~~~~~~~~~~~~\left[k^4(C^2_1+2C_0C_2)V_{i-4-j-m}-k^2\omega^2C_2(C_{i-4-j-m}+2V_{i-4-j-m})\right]\\
&+2k^4C_1C_2\sum_{m=0}^{i-5}\sum_{j=0}^{i-5-m}V_{i-5-j-m}(m+2)(m+1)(2C_j+\gamma V_j)\chi_{m+2}\\
&+k^4C^2_2\sum_{m=0}^{i-6}\sum_{j=0}^{i-6-m}V_{i-6-j-m}(m+2)(m+1)(2C_j+\gamma V_j)\chi_{m+2},
\end{aligned}
\end{equation}
\newpage
\begin{equation}
\begin{aligned}
&D_2(k,\omega^2)  \\
&=\sum_{m=0}^{i-2}\sum_{j=0}^{i-2-m}(j+1)(m+1)\chi_{m+1}\left[\omega^4(2-\gamma)(C_{i-2-j-m}V_{j+1}-C_{j+1}V_{i-2-j-m})
\right.\\
&~~~~~~~~~~~~~~~~+k^2\omega^2C_0(4V_{i-2-j-m}C_{j+1}+\gamma C_{i-2-j-m}V_{j+1}+2C_{i-2-j-m}C_{j+1}-4C_{i-2-j-m}V_{j+1})\\
&~~~~~~~~~~~~~~~~\left.+2k^4C^2_0(C_{i-2-j-m}V_{j+1}-V_{i-2-j-m}C_{j+1})\right]\\
&+\sum_{m=0}^{i-3}\sum_{j=0}^{i-3-m}(j+1)(m+1)\chi_{m+1}\left[4k^4C_0C_1(C_{i-3-j-m}V_{j+1}-V_{i-3-j-m}C_{j+1})\right.\\
&~~~~~~~~~~~~~~~~+\left.k^2\omega^2C_1(4V_{i-3-j-m}C_{j+1}+\gamma C_{i-3-j-m}V_{j+1}+2C_{i-3-j-m}C_{j+1}-4C_{i-3-j-m}V_{j+1})\right]\\
&+\sum_{m=0}^{i-4}\sum_{j=0}^{i-4-m}(j+1)(m+1)\chi_{m+1}\left[2k^4(C^2_1+2C_0C_2)(C_{i-4-j-m}V_{j+1}-V_{i-4-j-m}C_{j+1})\right.\\
&~~~~~~~~~~~~~~~~\left.+k^2\omega^2C_2(4V_{i-4-j-m}C_{j+1}+\gamma C_{i-4-j-m}V_{j+1}+2C_{i-4-j-m}C_{j+1}-4C_{i-4-j-m}V_{j+1})\right]\\
&+4k^4C_1 C_2\sum_{m=0}^{i-5}\sum_{j=0}^{i-5-m}(j+1)(m+1)\chi_{m+1}(C_{i-5-j-m}V_{j+1}-V_{i-5-j-m}C_{j+1})\\
&+2k^4C^2_2\sum_{m=0}^{i-6}\sum_{j=0}^{i-6-m}(j+1)(m+1)\chi_{m+1}(C_{i-6-j-m}V_{j+1}-V_{i-6-j-m}C_{j+1}),
\end{aligned}
\end{equation}
and
\begin{equation}
\begin{aligned}
%&D_3(k,\omega^2)	\\
D_3(k,\omega^2)&=\omega^6\sum_{m=0}^{i-2}(2C_{i-2-m}+\gamma V_{i-2-m})\chi_m\\
&+\sum_{m=0}^{i-2}\sum_{j=0}^{i-2-m}\chi_m \left[-k^2\omega^4(2C_j+\gamma V_j)(2C_{i-2-j-m}+V_{i-2-j-m})\right.\\
&~~~~~~~~~~~~~~~~+k^4\omega^2C_0(2C_j+\gamma V_j)(C_{i-2-j-m}+2 V_{i-2-j-m})\\
&~~~~~~~~~~~~~~~~\left.-k^6C^2_0(2C_j+\gamma V_j)V_{i-2-j-m}\right]\\
&+\sum_{m=0}^{i-3}\sum_{j=0}^{i-3-m}\chi_m\left[k^4\omega^2C_1(2C_j+\gamma V_j)(C_{i-3-j-m}+2 V_{i-3-j-m})\right.\\
&~~~~~~~~~~~~~~~~\left.-2k^6C_0C_1(2C_j+\gamma V_j)V_{i-3-j-m}\right]\\
&+\sum_{m=0}^{i-4}\sum_{j=0}^{i-4-m}\chi_m\big\{k^4\omega^2C_2(2C_j+\gamma V_j)(C_{i-4-j-m}+2 V_{i-4-j-m})\\
&~~~~~~~~~~~~~~~~-k^6(C^2_1+2C_0C_2)(2C_j+\gamma V_j)V_{i-4-j-m}\big\}\\
&-2 k^6C_1C_2\sum_{m=0}^{i-5}\sum_{j=0}^{i-5-m}(2C_j+\gamma V_j)V_{i-5-j-m}\chi_m\\
&-k^6C^2_2\sum_{m=0}^{i-6}\sum_{j=0}^{i-6-m}(2C_j+\gamma V_j)V_{i-6-j-m}\chi_m.
\end{aligned}
\end{equation}

 \clearpage

\clearpage

\clearpage

\clearpage

\clearpage

\end{document}